\documentstyle[prd,aps,preprint]{revtex}
\begin{document}
\draft

%
%

\preprint{Nisho-98/3} \title{Josephson Effects in Double-Layer Quantum Hall
States} \author{Aiichi Iwazaki}
\address{Department of Physics, Nishogakusha University, Shonan Ohi Chiba
  277,\ Japan.} \date{April 15, 1998} \maketitle
\begin{abstract}
Under quite plausible assumptions on double-layer quantum Hall states
with strong interlayer correlation, we show in general framwork that 
coherent tunneling of a single electron between two layers is possible.
It yields Josephson effects with unit charge tunneling. The origin is that 
Halperin states in the quantum Hall states are highly degenerate
in electron number difference between two layers in the absence of 
electrons tunneling. 
\end{abstract}
\vskip2pc

It has been shown\cite{ei,wz} that in double-layer ( DL )
quantum Hall states with filling factor, $\nu=1$, a zero energy
mode arises owing to spontaneous symmetry breakdown of pseudo-spin 
U(1) symmetry. This symmetry is associated with the conservation of electron 
number difference between two layers. Thus the mode arises only for 
the system with no interlayer-tunneling; the tunneling leads obviously 
to non-conservation of the electron number difference and so breaks 
the symmetry explicitly. In Chern-Simons gauge theory this symmetry 
breakdown is realized\cite{ei,ei2} 
as a condensation of bosonized electrons; this is 
quite similar to the condensation of Cooper pairs in superconductors. 
Thus one working
with the theory is led naturally to anticipate the existence of Josephson-like
effects in the DL quantum Hall states. Actually it has been 
argued\cite{ei,ei2,wz2,ei3} by 
addressing the mode that Josephson effects may arise in these DL quantum 
Hall states. It has, however, been pointed out\cite{m} 
that the interlayer-tunneling
gives a gap to the mode and freezes a phase degree of freedom 
whose existence is essential for Josephson effects. But this gap is 
owing to so-called Anderson plasmon\cite{ei3} and 
is a universal property of standard
Josephson junctions. Thus the phase coherence between two layers is still
alive and Josephson effects are expected in the system.

In this paper using quite plausible assumptions we show 
that Josephson effects are possible in the DL quantum Hall states with $\nu=1$.
The assumptions are that in the case of no interlayer-tunneling, quantum Hall
states in the DL system are described with Halperin states\cite{h} 
and that such Halperin states are degenerate in 
the electron number difference between two layers, when small capacitance 
energies are neglected. Furthermore, we assume that an interaction describing
interlayer tunneling conserves angular momentum ( $J$ ) of electron; an 
electron with $J=m$ in a layer tunnels into a state with $J=m$ in the 
other layer. It is also assumed that the energy scale of the tunneling is 
much smaller than the Coulomb energy, $e^2/l_B$ where $l_B$ is the magnetic 
length and $-e$ is the charge of electron.

In the assumptions the degeneracy of Halperin states in the electron number
difference is the consequence of the broken U(1) symmetry 
in the system with no interlayer-tunneling. Hence it seems to be 
quite acceptable. The assumption of angular momentum conservation is hold
in the system with no irregularity which breaks the rotational symmetry.
The last assumption on the energy scale guarantees that the tunneling effect
can be treated perturbatively. Later we argue that possible deviations 
from these assumptions in realistic samples do not affect seriously the 
Josephson effects.

As has been shown numerically\cite{n} 
in the $\nu=1$ DL system with an appropriate
interlayer distance, $d$, comparable to $l_B=\sqrt{1/eB}$ ( $B$ is magnetic 
field perpendicular to the layers ), 
Halperin states are fairly
good approximate ground states in the case of no interlayer-tunneling.
Even in the presence of  the interlayer-tunneling, 
the ground state may be approximated by mixture of the states. 

First, we briefly sketch Halperin states with the filling factor $\nu=1/m$
($m$ = integer).
They are described as

\begin{equation}
\Psi(N_1,N_2)=\prod_{l<k}(z_l-z_k)^m\prod_{s<t}(w_s-w_t)^m
\prod_{l,s}(z_l-w_s)^m\exp(-\sum(|z|^2+|w|^2)/4l_B^2)
\end{equation}
where $N_1$ and $N_2$ are the numbers of electrons in each layer. We denote
complex coordinates of electrons in each layer 
with $z$ and $w$, respectively. These states are degenerate with each other 
in the electron number difference, $N_1-N_2$, when we neglect 
charging energies. Namely, when the filling factor $\nu=1/m$ is given, 
total number of electrons, $N_1+N_2$ is determined uniquely but the values of
$N_1-N_2$ are arbitrary. This means that there are many quantum Hall 
states with $\nu=1$,
which are characterized by quantum numbers, $N_1-N_2$. 
They are all degenerate.
Once we take account of the charging energies, 

\begin{equation}
H_c=\frac{e^2\hat{S_z}^2}{C}
\end{equation}
the degeneracy is lifted up, where $C=\epsilon L^2/4\pi d$ is the electric
capacitance of the double-layers; $L^2$ is the surface area of the layers 
and $\epsilon$ is the dielectric constant between two layers.
The operator $\hat{S_z}$ represents electron number difference and its eigen
value is $(N_1-N_2)/2$. Electric neutrality is assumed for the state with
$N_1-N_2=0$. Obviously the charging energy is small 
enough for infinitely large $L^2$. Hence the Halperin states are almost
degenerate even if the charging energy is switched on.

In the realistic DL samples, 
the states with $\nu=1$ have been realized\cite{ma,e}.
Hence we only consider such states. First we note that 
in each layer, one particle states in the Lowest Landau level are 
characterized by the angular momentum, $J=m$ ( $m$ takes a value
of $1$ through $(N_1+N_2)/2$ ). Then,
Halperin states with $\nu=1$ are states such that
either of electrons in the first or second
layer occupies each one particle eigenstate with $J=m$, but electrons 
in both layers do not occupy simultaneously the states 
with the same angular momentum.
More explicitly, the state can be written such that 

\begin{equation}
|Halperin>=P(N_1,N_2)\prod_{m=0}^{N_1+N_2-1}(a_m^{\dagger}+b_m^{\dagger})|0>
\end{equation}
with $a_m|0>=b_m|0>=0$,
where $P(N_1,N_2)$ is the projection operator picking up a state with
$N_1$ electrons on the 1st layer and $N_2$ electrons on the 2nd layer.
$a_m$ and $b_m$ are annihilation operators of electrons with angular
momentum, $m$, on the 1st layer and 2nd layer respectively.

When the interlayer-tunneling is allowed, these Halperin states are 
mixed with each other. We assume the following interlayer-tunneling
interaction, 

\begin{equation}
H_t=-\frac{\Delta_{sas}}{2}\int (\Psi_1^{\dagger}\Psi_2+h.c.)=
-\frac{\Delta_{sas}}{2}(\sum_m a_m^{\dagger}b_m+h.c.+ 
higher\\\ Landau\\\ levels )
\end{equation}
where $\Psi_i$ is the electron field of i-th layer and $\Delta_{sas}$
is the energy difference of symmetric ( $\Psi_1+\Psi_2$ ) and 
antisymmetric ( $\Psi_1-\Psi_2$ ) states. This tunneling term 
preserves the angular momentum of electrons.

It is interesting to see that a Halperin state, $|S,S_z>$, 
which is an eigenstate of $\hat{S_z}$ with the eigenvalue of $(N_1-N_2)/2$
( $S=(N_1+N_2)/2$ ),
is transformed to the state, $|S,S_z\pm1>$ by the tunneling interaction,

\begin{equation}
H_t|S,S_z>=-\frac{\Delta_{sas}}{2}(\sqrt{(S-S_z)(S+S_z+1)}|S,S_z>+
\sqrt{(S+S_z)(S-S_z+1)}|S,S_z-1>).
\end{equation}
It can be also derived by using $O(3)$ algebra, $[\hat{S_i},\hat{S_j}]
=i\epsilon_{ijk}\hat{S_k}$, where

\begin{equation}
\hat{S_x}=\frac{1}{2}\int(\Psi_1^\dagger\Psi_2+h.c.),\\\
\hat{S_y}=\frac{-i}{2}\int(\Psi_1^\dagger\Psi_2-\Psi_2^\dagger\Psi_1),\\\
and \\\ \hat{S_z}=\frac{1}{2}\int(\Psi_1^\dagger\Psi_1-\Psi_2^\dagger\Psi_2).
\end{equation}

In this notation $H_t=-\Delta_{sas}\hat{S_x}$.
These operators, $\hat{S_i}$, are called pseudospin operators and 
Halperin states, $|S,S_z>$ are elements of a representation space 
of the operators. Namely they represent the states with the pseudospin, $S$,
whose $z$ component is $S_z$.

As far as we neglect the effects of the charging energy and the tunneling  
interaction, Halperin states $|S,S_z>$ are realized as 
quantum Hall states with $\nu=1$. 
These states are degenerate in the quantum number, $S_z$.
But once we include these effects, the degeneracy is lifted up. 
To find the groundstate, we need to diagonalize the Hamiltonian,

\begin{equation}
H=H_c + H_t=\frac{e^2\hat{S_z}^2}{2C}-\Delta_{sas}\hat{S_x}
\label{H}
\end{equation} 
in the space of Halperin states, $|S,S_z>$.
Thus we obtain an eigenstate of the Hamiltonian,

\begin{equation}
|G>=\sum_n A_n|S,n> 
\end{equation}
where $A_n$ satisfies the recursion formula,

\begin{equation}
EA_n=\frac{e^2n^2}{2C}A_n-\frac{\Delta_{sas}}{2}
[A_{n-1}\sqrt{(S-n+1)(S+n)}+A_{n+1}\sqrt{(S+n+1)(S-n)}],
\end{equation}
with $E$ being the energy of the eigenstate $|G>$.
In order to solve the recursion formula, 
we assume $S=(N_1+N_2)/2$ being much
larger than any $n$, in other words, $A_n$ with $n$ being the order of S
is small enough to be neglected. Then we expand the square root, leaving
the terms of the lowest order in $n/S$. Setting 
$\Psi(\theta)=\sum A_ne^{i\theta n}$, we rewrite the formula such that

\begin{equation}
E\Psi(\theta)=-\frac{e^2}{2C}\frac{\partial^2}{\partial\theta^2}\Psi(\theta)
-S\Delta_{sas}\cos\theta\Psi(\theta).
\end{equation}

This $\Psi(\theta)$ represents the wave function of the eigenstate $|G>$
in terms of angle variable $\theta$ conjugate to the electron number
difference $n\sim N_1-N_2$ between two layers,

\begin{equation}
\Psi(\theta)=<\theta|G>, 
\end{equation}
with $|\theta>=\sum_n e^{-i\theta n}|S,n>$

A solution of this Shr$\ddot{o}$dinger-like equation can be obtained 
by assuming a particle 
sitting in the bottom of the cosine potential, i.e. $\theta\ll 1$,
 
\begin{equation}
\Psi(\theta)\sim \exp(-mE_0\theta^2/2)
\label{p}
\end{equation}
where $m=C/e^2$ is the mass of the particle and 
$E_0=\sqrt{e^2S\Delta_{sas}/C}$ is 
the energy of the state. 

Here we comment that since $\theta$ is the conjugate variable
to $S_z\sim N_1-N_2$, 
roughly it represents a direction of the pseudospin in x-y plane.
Thus the above solution represents a state 
with the direction of the pseudospin pointed to x axis, i.e. $\theta=0$. 
This fact
can be understood by noting the existence of the term, $\Delta_{sas}\hat{S_x}$ 
in the Hamiltonian; the term implies the imposition of magnetic field,
$B=\Delta_{sas}$ pointed to x axis. Thus the spin is pointed to x axis.

We also comment that since the fluctuation of $\theta$ is given by 
$(mE_0)^{-1/2}\propto \Delta_{sas}^{1/2}$, it diverges in the limit of 
a vanishing tunneling amplitude $\sim \Delta_{sas}$ of a electron. Thus 
the phase $\theta$ is not well defined and Josephson 
phenomena are not expected to be seen in the limit just as in superconducting 
Josephson junctions.


As can be seen easily, there exist solutions representing the pseudospin
rotating around z axis; the particle is not bounded to 
the cosine potential and it moves from $\theta=-\infty$ to $\theta=+\infty$.
Obviously, this mode does not correspond to the groundstate of the 
system eq(\ref{H}). But the mode is excited by applying a voltage, $V_0$ 
between two layers. To see it, we add a term, $eV_0\hat{S_z}$ to the 
Hamiltonian,

\begin{equation}
H'=H + eV_0\hat{S_z}=\frac{e^2\hat{S_z}^2}{2C}-
\Delta_{sas}\hat{S_x}+eV_0\hat{S_z} 
\label{h}
\end{equation} 

Then, it follows with the similar manipulation to the above one that 
the wave function $\Psi(\theta)$ satisfies the following equation,

\begin{equation}
(i\partial_t + ieV_0\partial_{\theta})\Psi(\theta)=
-\frac{e^2}{2C}\partial_{\theta}^2\Psi(\theta)-S\Delta_{sas}\cos\Psi(\theta),
\label{s}
\end{equation}
where we have explicitly indicated the derivative in time, $t$, in order to 
see the evolution of $\Psi$. Assuming a wave packet, $\Psi(\theta)$,
in $\theta$, we can show that 

\begin{equation}
\frac{d}{dt}<\theta>=\frac{d}{dt}\int\overline{\Psi}\theta\Psi d\theta=
eV_0+\frac{e}{C}<en>=eV_0 + eV_{ind}\equiv eV_{me},
\end{equation}
where $V_{ind}=e<n>$ denotes an induced voltage associated with the charging,
$<en>$ and $V_{me}$ does the voltage measured actually.

This implies that the phase, $\theta$, conjugate to the electron
number difference, $N_1-N_2$, evolves according to the standard Josephson
equation. In terms of the pseudospin languages, the spin rotates around z axis.
Furthermore, we can show that the tunneling current is given by 

\begin{equation}
\frac{d}{dt}<-en>=\frac{d}{dt}\int\overline{\Psi}
ei\frac{\partial}{\partial\theta}\Psi d\theta=eS\Delta_{sas}<\cos\theta>
\approx J_c\cos(<\theta>),
\label{J}
\end{equation} 
with $J_c=eS\Delta_{sas}$,
where $J_c$ is the critical current. Hence we can see that the current 
is the same as the one in standard Josephson effects; conventionally 
the current is given by $J_c\sin\theta$, which is obtained by
shifting $\theta\to \theta+\pi/2$ in eq(\ref{J}). 

Consequently, we obtain the Josephson equations controlling the phase 
and the tunneling current in the quantum Hall system.

It is instructive to see the quantum Hall 
state Josephson effects in a different way.
We may rewrite Hamiltonian eq(\ref{h}) or equivalently Hamiltonian read 
from Schr$\ddot{o}$dinger equation eq(\ref{s}) as follows,

\begin{equation}
H'=\frac{p^2}{2m}+eV_0p-S\Delta_{sas}\cos\theta,
\end{equation}
with $p=-i\frac{\partial}{\partial\theta}$.

Using this Hamiltonian we find that the velocity of $\theta$ is 
given by $\dot{\theta}=p/m+eV_0$ where a dot denotes a time derivative.
On the other hand a time variation of $p$ is given by 
$\dot{p}=-S\Delta_{sas}\sin\theta$. Thus it follows that 
$m\ddot{\theta}=-S\Delta_{sas}\sin\theta$. This is the equation of pendulum.
The only effect of 
switching on the voltage, $V_0$, changes the velocity of the pendulum 
by $eV_0$. Before applying the voltage its momentum $p$ takes a value 
of the order of $\sqrt{mE_0}$ in the bottom of the cosine potential 
as shown in eq(\ref{p}). Hence the velocity, 
$\dot{\theta}\sim \sqrt{E_0/m}\sim \sqrt{1/L^2}$, is quite small.
However, once the voltage is switched on, the pendulum gains a velocity,
$eV_0$, and so it can climb the mountain of the potential, when $eV_0$
is sufficiently large. Thus the pendulum can rotate
as $\theta$ increases without limit; the pseudospin 
can rotate around z axis. In other words, Josephson effects arise owing 
to the voltage between two layers.

The essence of this phenomena, coherent interlayer tunneling 
of a single electron, is that Halperin states, $|S,n>$,
are highly degenerate 
in their quantum number $n=(N_1-N_2)/2$. Namely 
even if electrons move from a layer to the other one, the energy of 
the system does not change.
The situation is quite similar to 
that of Josephson junctions, in which states composed of two superconducting
states are degenerate in the difference of numbers of Cooper pairs involved
in each superconducting state. That is, 
the energy of the system does not change 
even if Cooper pairs move from a superconductor to the other one.  
In both cases this degeneracy leads to 
the existence of the phase conjugate to the 
quantum number $n$; the phase becomes a good quantum number characterizing
the quantum Hall state as well as the state of Josephson junction.

Only difference between the case of the quantum Hall states and 
that of Josephson junction is that in the 
DL quantum Hall states, there does not exist a phase degree 
of freedom associated with each state of the two layers, on the other hand 
there exist a phase degree of freedom 
associated with each state of two superconductors in
Josephson junction.

As we have shown under the plausible and general assumptions, 
Josephson effects arise in the DL quantum Hall states with $\nu=1$.
Some of these assumptions have been confirmed numerically. In the realistic 
samples however these purely assumptions do not necessarily hold. But small 
deviations from the above hypothetic Hamiltonian or deviations from 
the Halperin states may be regarded as effects of impurities. These effects 
may not change our results because coherent phenomena like Josephson effects 
are not seriously affected by the impurities in general.
Therefore we expect that the effects arise in the DL quantum Hall states.

Recent experiment\cite{e} on tunneling currents between two layers ( whose 
separations are much larger than ones claimed 
for the existence of Josephson effects ) shows the existence of an exciton 
made of a tunneled electron and a hole it leaves behind; the energy of 
the exciton yields actual tunneling barrier. It suggests a state of 
the exciton with zero energy when the interlayer separation is 
sufficiently small, but nonvanishing. Then the tunneling barrier is 
expected to vanish for the system and electrons tunnel freely without 
friction. This may reads to Josephson effects in the DL quantum Hall states,
as we have discussed in this paper.

The author would like to express thanks to members of 
particle theory group for their hospitality in 
tanashi, KEK.





\end{document}